\documentclass[12pt,a4paper]{article}

\usepackage[margin=1.8cm]{geometry}
\usepackage{graphicx}
\usepackage{amsmath}
\usepackage{amssymb}
\usepackage{bm}
\usepackage{hyperref}
\usepackage{xcolor}
\colorlet{RED}{red}
\usepackage{mathtools}
\usepackage{float}
\usepackage[font=small,labelfont=bf]{caption}
\usepackage{authblk}
\usepackage{setspace}
\usepackage{titlesec}

\titleformat{\section}{\large\bfseries\centering}{\Roman{section}.}{0.5em}{\MakeUppercase}
\titleformat{\subsection}{\normalsize\bfseries\itshape}{\Alph{subsection}.}{0.3em}{}
\titleformat{\subsubsection}[runin]{\normalsize\itshape}{}{0em}{}[.]

\newcommand{\vF}{v_{\mathrm{F}}}
\newcommand{\vx}{v_x}
\newcommand{\vy}{v_y}
\newcommand{\wt}{w}
\newcommand{\wx}{w_x}
\newcommand{\wy}{w_y}
\newcommand{\kvec}{\mathbf{k}}
\newcommand{\rvec}{\mathbf{r}}
\newcommand{\kpar}{k_{\parallel}}
\newcommand{\kperp}{k_{\perp}}
\newcommand{\deff}{d_{\mathrm{eff}}}

\newcommand{\xeff}{x_{\mathrm{eff}}}
\newcommand{\EF}{E_{\mathrm{F}}}

\begin{document}

\title{\Large\textbf{All-Electrostatic Valley Filtering by Barrier Rotation in Tilted Dirac/Weyl Semimetals}}

\author[1]{Can Yesilyurt\thanks{canyesil@nanorescenter.com; can{-}{-}yesilyurt@hotmail.com}}
\affil[1]{Nanoelectronics Research Center, Istanbul, Turkey}

\date{\today}

\maketitle

\begin{abstract}
\noindent
Charge carriers in Dirac/Weyl semimetals with tilted anisotropic energy dispersion exhibit valley-dependent refraction and reflection at electrostatic barrier interfaces. Here, we show that an angled barrier interface provides a purely electrostatic route to valley filtering, producing finite valley-polarized conductance. We develop a generalized transfer-matrix formalism for the tilted, anisotropic Dirac Hamiltonian, extend it to treat electrostatic barriers at arbitrary angles, and calculate the transmission in the rotated-barrier frame. We also present simulated valley-resolved trajectories in a finite device geometry, demonstrating that one valley is selectively transmitted, while the other is predominantly reflected by the angled barrier, without invoking real or pseudomagnetic fields.
\end{abstract}

\section{Introduction}

\label{sec:intro}

Charge carriers in two-dimensional (2D) Dirac materials may carry a valley isospin degree of freedom, in addition to their real spin and charge~\cite{Rycerz2007,Xiao2007,Schaibley2016}. The valley index labels the degenerate but inequivalent $K$ and $K'$ Dirac points in momentum space and has emerged as a promising carrier of information for \emph{valleytronics}~\cite{Rycerz2007,Schaibley2016}. Generating and controlling valley-polarized currents requires breaking the degeneracy between the two valleys and then selectively filtering the transmission of one valley over the other~\cite{Rycerz2007,Xiao2007}. A considerable amount of effort has been devoted to achieving valley-dependent tunneling in condensed-matter systems. Valley polarization has been demonstrated by using magnetic fields~\cite{Settnes2016,Williams2011,Taychatanapat2015}, optical helicity~\cite{Mak2012}, and line defects~\cite{Rycerz2007,Gunlycke2011}. One possible avenue is to apply uniaxial strain to induce a relative shift of the two valleys in momentum space; as a result, electrons in different valleys refract in opposite directions at the interface between strained and unstrained regions~\cite{Pereira2009}. Although the gradient of uniaxial strain results in the angular separation of electron trajectories according to the valley index, this alone does not give rise to valley-polarized conductance since the contributions of both valleys to the overall conductance remain identical. To induce valley-polarized transport, one must additionally break the angular symmetry of the transmission profile; one established route is the transverse Lorentz displacement produced by a magnetic barrier ~\cite{Fujita2010,Yesilyurt2015,Yesilyurt2016AIP,Yesilyurt2016SciRep,Zhai2010}. However, each of these approaches presents practical limitations. External magnetic fields complicate on-chip integration, ferromagnetic layers increase fabrication complexity and limit tunability, strain engineering requires nanoscale mechanical control, and line defects demand atomic precision. Additionally, modulating valley polarization would require either altering the direction of the applied strain or switching the magnetization direction. More generally, such schemes first separate the two valleys and then extract a net polarization through a second, magnetic-, strain-, or magneto-electric-based stage~\cite{Yesilyurt2019,Li2018,Zheng2021,ZhangTVHE2023}.

A qualitatively different avenue has opened with the discovery of tilted anisotropic materials~\cite{Goerbig2008,OBrien2016}. This broad and growing class of tilted Dirac/Weyl materials includes transition metal tellurides (e.g., WTe$_2$) ~\cite{Soluyanov2015,Ali2014}, the predicted 8-\textit{Pmmn} borophene~\cite{Lopez2016,Zabolotskiy2016}, the organic conductor $\alpha$-(BEDT-TTF)$_2$I$_3$~\cite{Goerbig2008}, and strained graphene variants~\cite{Goerbig2008}.  In these systems, the energy dispersion acquires a tilt term that gives rise to a transverse momentum shift along the tilt direction in the presence of an electrical potential gradient~\cite{Yesilyurt2017APL, Nguyen2018}. It has previously been shown that this chirality-dependent refraction at the barrier interface polarizes the Weyl fermions in angle-space according to their valley index~\cite{Yesilyurt2019}. In addition, the perfect transmission angle shifts to a finite \emph{oblique Klein tunneling angle} $\theta_{\mathrm{K}}$, which is a material constant independent of energy and barrier height~\cite{Zhang2018,Kong2021,Nguyen2018}. Crucially, when the two valleys are tilted in opposite directions, the Klein tunneling peaks occur at $+\theta_{\mathrm{K}}$ for the $K$ valley and $-\theta_{\mathrm{K}}$ for $K'$, which breaks the valley degeneracy in angle-space~\cite{Zhang2018,Kong2021}. The tilt-induced valley splitting in angle-space, combined with an appropriate symmetry-breaking mechanism, can produce net valley polarization~\cite{Yesilyurt2019, Zheng2021, ZhangTVHE2023}. 
 
It has been shown that the interface orientation itself can strongly reshape the angular transmission in graphene, phosphorene, and $1T'$-MoS$_2$~\cite{Sajjad2012,BetancurOcampo2019,Wang2023}. A related study examined valley-dependent retroreflection and anomalous Klein tunneling in an ideal 8-\textit{Pmmn} borophene $n$-$p$-$n$ junction, with transport directed along the barrier normal and the transverse momentum conserved so that the junction angle enters only by rotating the anisotropic tilted Fermi surface relative to the fixed normal contacts~\cite{Zhou2019}. In contrast, practical devices have finite channel widths, which allow carriers reflected at barrier interfaces to scatter off channel walls and subsequently re-encounter the barrier at modified angles~\cite{Sajjad2012}. Furthermore, the influence of specific device geometry, such as angled potential barriers, as well as the combined effects of barrier angle, barrier height, and finite dimensions on filtering efficiency remain unaddressed. Therefore, in this work, we focus on this finite-width, angled-gate regime, in which the source--drain axis is kept fixed while the electrostatic top gate is tilted relative to it, so that $x_{\mathrm{eff}}=x-y\tan\alpha$ and $k_{\parallel}=k_x\sin\alpha+k_y\cos\alpha$; the conductance is then integrated over the source modes $k_y$ of the transport channel, and a finite-width trajectory calculation determines which of these modes reach the drain. Consequently, we show that an angled electrostatic gate on a tilted Dirac material can generate valley-polarized currents without the need for magnetic fields or strain engineering, with the tilted Dirac cone supplying the valley-split angular transmission while barrier rotation and the finite channel geometry supply the angular selection. We develop a generalized transfer-matrix method (TMM) formalism for the tilted anisotropic Dirac Hamiltonian, extended to treat barriers at arbitrary angles $\alpha$ relative to the transport direction. We further develop a semiclassical trajectory simulation in the full 2D device geometry and directly demonstrate that the angled gate selectively transmits one valley while reflecting the other, thereby enabling valley filtering rather than merely valley separation. We apply this framework to 8-\textit{Pmmn} borophene $n$-$p$-$n$ junctions as a concrete example and show that the barrier angle $\alpha$ converts the intrinsic tunneling asymmetry into a net valley-polarized conductance, which is gate-tunable through the barrier height $V_0$. The mechanism is general and applies to any material that hosts tilted Dirac or Weyl fermions.

\section{Device geometry and model}

\label{sec:device_model}

The proposed valley filter device is shown in Fig.~1. The device consists of a tilted Dirac material channel (e.g., 8-\textit{Pmmn} borophene or few-layer WTe$_2$ flake) with source and drain contacts, a global back gate that sets the Fermi energy $\EF$, and a tilted top gate that defines an angled electrostatic barrier of height $V_0$, as illustrated in Fig.~1(a). The barrier region (shaded blue) is tilted by angle $\alpha$ relative to the $y$-axis, as shown in Fig.~1(c) and (d). The energy-band diagram is depicted in Fig.~1(b), where the tilted Dirac cones in the barrier region are shifted by $V_0$ relative to those in the free region.

The physical mechanism of the valley filter can be understood intuitively from the schematic diagrams in Fig.~1(c) and (d). For a straight barrier ($\alpha = 0$), as shown in Fig.~1(c), the $K$ (blue, dotted) and $K'$ (red, dotted) carriers refract symmetrically upon entering the barrier, with the oblique Klein tunneling angle $\theta_{\mathrm{K}}$ splitting them in opposite transverse directions. However, this angular separation does not produce net valley polarization in the integrated conductance since the transmission profile is sampled symmetrically over all incident angles. When the barrier is tilted by $\alpha = 20^\circ$, as shown in Fig.~1(d), the geometry breaks this symmetry: one valley (here $K$, blue) encounters the barrier near its oblique Klein tunneling peak and transmits along a nearly straight trajectory, whereas the other valley ($K'$, red) approaches at unfavorable angles and is therefore reflected from the barrier interface.

\begin{figure}[H]
\centering
\includegraphics[width=0.95\textwidth]{Figure-1.pdf}
\caption{\label{fig:device}\textbf{Proposed valley filter device and operating principle.} (a)~Device schematic with source and drain contacts, a global back gate, and a tilted top gate defining the angled barrier of height $V_0$. (b)~Schematic energy-band diagram showing the transversely tilted Dirac cones in each region; the barrier-region cones are shifted by $V_0$, and the Fermi energy $\EF$ is horizontal. Shaded regions denote occupied states. (c)~Top-view illustration of the device with a straight barrier ($\alpha = 0$): $K$ (blue, dotted) and $K'$ (red, dotted) carriers refract symmetrically at the oblique Klein tunneling angle $\theta_{\mathrm{K}}$, producing no net valley polarization in the integrated conductance. (d)~Tilted barrier ($\alpha = 20^\circ$): the angled barrier selectively transmits one valley while reflecting the other, giving rise to valley-polarized conductance.}
\end{figure}

\subsection{Hamiltonian and energy dispersion}
\label{sec:hamiltonian}

The low-energy quasiparticles near a single Dirac or Weyl point in a tilted anisotropic 2D material can be described by a general low-energy Hamiltonian~\cite{Goerbig2008,Zabolotskiy2016,Kong2021}
\begin{equation}\label{eq:H}
H = \hbar \vx \sigma_x k_x + \hbar \vy \sigma_y k_y + \hbar s(\wx k_x + \wy k_y)\mathbb{I} + V(\rvec)\mathbb{I},
\end{equation}
where $\sigma_{x,y}$ are Pauli matrices acting on the sublattice pseudospin, $V(\rvec)$ is the external electrostatic potential, $\vx$ and $\vy$ are anisotropic Fermi velocities along the transport ($x$) and transverse ($y$) directions, and $\wx$ and $\wy$ are tilt velocity components. The sign $s = \pm 1$ carries the valley index; in many tilted Dirac/Weyl materials, the two inequivalent valleys are tilted in opposite directions due to the underlying crystal symmetry (often linked to broken inversion symmetry), but this is not a universal rule, and multi-valley systems can host more complex tilt patterns. The type-I regime (i.e., closed Fermi surface) requires that the tilt magnitude remain smaller than the cone velocities; for a general tilt direction, this corresponds to $\sqrt{w_x^2/v_x^2 + w_y^2/v_y^2} < 1$. Solving the Hamiltonian in Eq.~\eqref{eq:H}, the eigenvalues are found as

\begin{equation}\label{eq:dispersion}
E_\lambda(\kvec) = V + \hbar s(\wx k_x + \wy k_y) + \lambda\sqrt{(\hbar\vx k_x)^2 + (\hbar\vy k_y)^2},
\end{equation}

with $\lambda = \pm 1$ the band index. Note that the pseudospin texture is determined solely by the anisotropic Fermi velocities $\vx$ and $\vy$, while the tilt enters only through the identity matrix. This decoupling means the tilt rigidly displaces the Fermi contour without altering the chiral structure~\cite{Nguyen2018}. For compactness we report velocity parameters in units of $\vF = 10^6$~m~s$^{-1}$ (e.g., $\vx = 0.86\,\vF$), while $\vx$, $\vy$, $\wx$, and $\wy$ remain dimensional velocities. For the numerical calculations presented below, we use parameters representative of 8-\textit{Pmmn} borophene: $\vx/\vF = 0.86$, $\vy/\vF = 0.69$, with a transverse tilt of magnitude $|\wt|/\vF = 0.32$ directed along the $y$-axis ($\phi_{\mathrm{t}} = 90^\circ$), giving a tilt ratio $\eta^2 = w_y^2/\vy^2 \approx 0.215$ (here $w_x = 0$, so the tilt is purely transverse)~\cite{Lopez2016,Zabolotskiy2016,Kong2021}. 8-\textit{Pmmn} borophene is a predicted type-I tilted Dirac material whose anisotropic properties are well established and widely used in the literature~\cite{Lopez2016}; these values serve as a representative parameter set. The Hamiltonian in Eq.~\eqref{eq:H} is general: the valley filtering mechanism is not restricted to borophene but applies to any two-dimensional or few-layer type-I tilted Dirac/Weyl system, yielding a finite valley-polarized conductance once the barrier angle and the crystal tilt axis are chosen appropriately with respect to the transport direction.

\subsection{Barrier geometry and conserved momentum}

For a straight barrier ($\alpha = 0$), translational invariance along the $y$-direction makes $k_y$ a conserved quantity, and the transverse tilt leads to the perfect-transmission direction shifting by $+\theta_{\mathrm{K}}$ for the $K$ valley and $-\theta_{\mathrm{K}}$ for $K'$, so that the valley degeneracy is broken in angle-space~\cite{Zhang2018,Kong2021,Nguyen2018,Zhou2019}.

We consider a one-dimensional rectangular potential barrier rotated by angle $\alpha$ relative to the $y$-axis, as illustrated in Fig.~1(c) and (d), where the conserved quantum number in the barrier frame is $\kpar = k_x\sin\alpha + k_y\cos\alpha$. The barrier potential is described by $V(x,y) = V_0\Theta(\xeff)\Theta(d - \xeff)$, where $\xeff = x - y\tan\alpha$ is the coordinate normal to the tilted barrier up to the scale factor $\cos\alpha$, $d$ is the barrier width measured along $x$ at $y=0$, and $\Theta$ is the Heaviside step function. The effective barrier width perpendicular to the interface is $\deff = d\cos\alpha$, which determines the transfer-matrix phase accumulation. The gradient of this potential produces a transverse force $F_y = -\partial V/\partial y = \tan\alpha\,\mathrm{d}V/\mathrm{d}\xeff$ at each interface, which is absent for a straight barrier. In the semiclassical picture, this force imparts a transverse momentum shift $\Delta k_y \propto \tan\alpha$ at each barrier interface. Here, $\alpha$ denotes the rotation of the barrier interface away from the $y$-axis (equivalently, the rotation of the barrier normal away from the source--drain, or transport, $x$-direction), not a rotation of the crystal axes: the leads inject modes along the fixed transport axis while the interface matching is performed in $\kpar$, so changing $\alpha$ changes how the injected modes sample the valley-asymmetric angular transmission profile.

\section{Results}
\label{sec:results}

\subsection{Valley-dependent conductance}
\label{sec:conductance_results}

The total valley-dependent conductance can be calculated considering the anisotropic band dispersion based on the Landauer--B\"uttiker formalism~\cite{Allain2011}. The valley-resolved ballistic conductance is given by
\begin{equation}\label{eq:G}
G_s = G_0\frac{W}{2\pi}\int_{k_y^-(s)}^{k_y^+(s)} T(E, k_y; s)\frac{v_x^{\mathrm{grp}}(k_y, s)}{\vF}\,\mathrm{d}k_y,
\end{equation}
where $G_0 = 2e^2/h$ is the spin-degenerate conductance quantum used for each valley-resolved channel, $W$ is the channel width, and $v_x^{\mathrm{grp}} = \wx s + \lambda v_x^2 k_x/\sqrt{(\vx k_x)^2 + (\vy k_y)^2}$ is the $x$-component of the anisotropic group velocity~\cite{Kong2021,Allain2011}. The integration limits $k_y^\pm(s)$ correspond to the boundaries of the propagating-mode window, which are determined by requiring $k_x$ to be real in the free region. For $\alpha \neq 0$, the transmission $T$ in Eq.~\eqref{eq:G} is evaluated using the conserved tangential momentum $\kpar$ from Sec.~\ref{sec:tmm_detail}, while the integration remains over the transverse modes $k_y$ of the free region (i.e., the incoming channel modes at the source contact).

To demonstrate the efficiency of the all-electrostatic valley-filter device, the proposed model is evaluated using material parameters for 8-\textit{Pmmn} borophene, and the conductance $G/G_0$ is calculated as a function of the electrostatic potential $V_0$. In Fig.~2(a), the conductance $G/G_0$ as a function of $V_0$ is shown for the straight-barrier ($\alpha = 0^\circ$) geometry illustrated in Fig.~2(e). The four curves correspond to the $K$ and $K'$ valleys obtained from both the TMM and the semiclassical approach. For each method, the $K$ and $K'$ curves coincide, confirming that a straight barrier produces no net valley polarization, which is consistent with previous theoretical predictions~\cite{Kong2021,Zhang2018}. The TMM and semiclassical results are in close agreement: in the $n$-$p$-$n$ regime ($V_0 > \EF$), the Fabry--P\'erot oscillations differ between the two methods because the finite channel width introduces wall reflections, as shown in the trajectory plots (see Fig.~3). This leads to trapped resonant modes at the barrier interfaces, which can either enhance or suppress conductance oscillations.

The valley-degenerate conductance profile changes qualitatively when the barrier is tilted. The conductance for $\alpha = 20^\circ$ is shown in Fig.~2(b) with the corresponding device geometry shown in Fig.~2(d). Below $V_0 \approx \EF$, the curves remain approximately degenerate. However, above $V_0 \approx \EF$, all four curves become clearly distinguishable: the $K$-valley conductance (both TMM and semiclassical) remains high, whereas the $K'$-valley conductance decreases substantially, as indicated by the arrows in Fig.~2(b). Within each valley, the TMM and semiclassical results track each other closely, with the semiclassical oscillations again somewhat smoothed by the finite device geometry. The splitting onset coincides with the transition from $n$-$n'$-$n$ to $n$-$p$-$n$ transport (i.e., when the barrier Dirac point crosses $\EF$), which is the regime where the Fermi surface mismatch between the two regions is more pronounced~\cite{Zhang2018,Kong2021}. Similarly, the valley-dependent conductance difference $\Delta G/G_0 = (G_K - G_{K'})/G_0$ is shown in Fig.~2(c), where it remains small (below approximately $0.5$) for $V_0 < \EF$ and increases to values on the order of  $\Delta G/G_0 \approx 3$--$4$ in the $n$-$p$-$n$ regime.

\begin{figure}[H]
\centering
\includegraphics[width=0.95\textwidth]{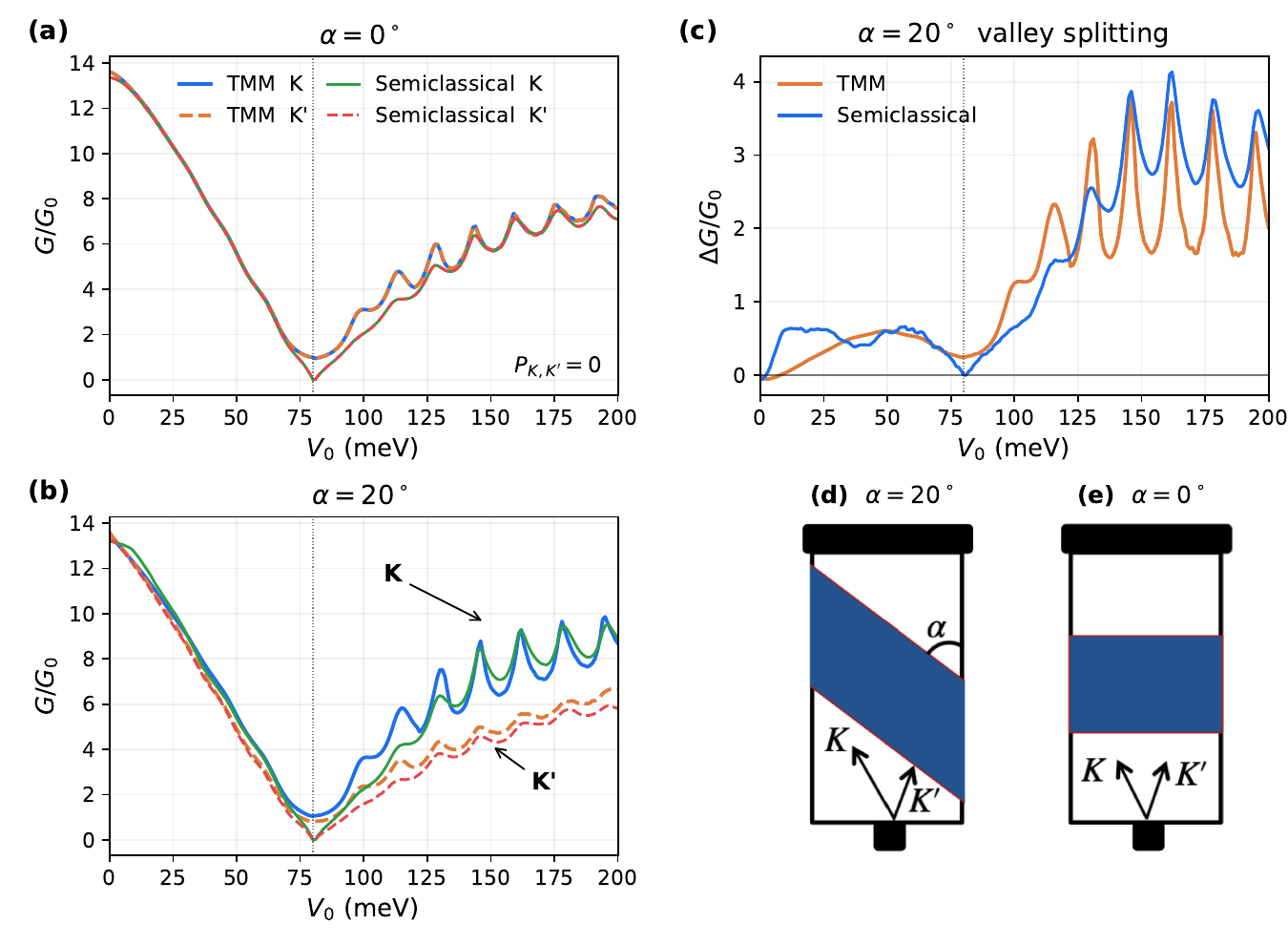}
\caption{\label{fig:conductance}\textbf{Valley-resolved conductance as a function of barrier height $V_0$.} (a)~Straight barrier ($\alpha = 0^\circ$), with the device geometry illustrated in (e): $G_K$ and $G_{K'}$ calculated by the transfer-matrix method (TMM; $K$: blue solid, $K'$: orange dashed) and the semiclassical approach ($K$: green solid, $K'$: red dashed), confirming zero valley polarization ($P = 0$). (b)~Tilted barrier ($\alpha = 20^\circ$), with the device geometry illustrated in (d): above $V_0 \approx \EF = 80$~meV, $K$ and $K'$ conductances split clearly, demonstrating the valley filtering effect. (c)~Valley-dependent conductance difference $\Delta G/G_0$ for $\alpha = 20^\circ$, showing the onset of valley polarization in the $n$-$p$-$n$ regime. (d),~(e)~Device schematics for the tilted and straight barrier configurations, respectively. Parameters: $\vx/\vF = 0.86$, $\vy/\vF = 0.69$, $|\wt|/\vF = 0.32$, $\phi_{\mathrm{t}} = 90^\circ$ (8-\textit{Pmmn} borophene), $W = 300$~nm, $L = 300$~nm, $d = 100$~nm, $\EF = 80$~meV.}
\end{figure}

\subsection{Valley-polarized carrier trajectories}

\label{sec:trajectories}

To demonstrate valley filtering in finite-device geometry, a semiclassical trajectory simulation is developed. While earlier works have shown that tilt produces valley-dependent refraction at a barrier interface~\cite{Yesilyurt2017APL,Yesilyurt2019,Nguyen2018}, the trajectory method developed here captures the full spatial dynamics and reveals that the angled barrier not only separates the two valleys in angle-space, but also actively filters them. To calculate the trajectories, at each barrier crossing, the parallel wave vector $\kpar$ is recorded, and the coherent Fabry--P\'erot transmission $T_{\mathrm{FP}}$ is evaluated from the sharp-barrier transfer matrix described in Sec.~\ref{sec:tmm_detail}. This hybrid approach uses the semiclassical trajectories to capture finite-geometry effects (i.e., wall reflections, angle remapping) while retaining the quantum-mechanical transmission amplitudes from the transfer matrix. The trajectory opacity is weighted by the transmission probability, enabling direct visualization of valley-resolved dynamics. The resultant conductance is given by
\begin{equation}\label{eq:G_traj}
G_s^{\mathrm{traj}} = G_0\frac{W}{2\pi}\sum_i T_{\mathrm{eff},i}(k_{y,i}; s)\frac{v_x^{\mathrm{grp}}(k_{y,i}, s)}{\vF}\Delta k_y,
\end{equation}
where $T_{\mathrm{eff},i}(k_{y,i}; s)$ is the coherent Fabry--P\'erot transmission evaluated at the $k_y$ of the $i$-th trajectory, and $\Delta k_y = (k_y^+ - k_y^-)/N$ is the uniform spacing over the propagating mode window of valley $s$. This discretization resembles the quantum Landauer integral in Eq.~\eqref{eq:G}, with the transfer matrix transmission replaced by the trajectory-assigned transmission. This semiclassical approach provides a spatially resolved picture of the valley filtering mechanism. The hybrid scheme retains the Fabry--P\'erot interference of each barrier traversal through the TMM weight, but it does not capture coherent interference between distinct real-space paths or diffraction from finite contacts and edges.
 The carrier trajectories for both valleys across three configurations are shown in Fig.~3, using the same device geometry ($W = 300$~nm, $L = 300$~nm) and transport parameters ($\EF = 80$~meV, $V_0 = 160$~meV). Blue trajectories correspond to the $K$ valley ($s = +1$) and red trajectories correspond to $K'$ ($s = -1$); the trajectory opacity is proportional to the transmission probability assigned at each barrier crossing. The barrier boundaries are indicated by black horizontal (or tilted) lines.

In graphene with $\alpha = 0^\circ$ [Fig.~3(a)], the $K$ and $K'$ trajectories are indistinguishable, which is a direct consequence of the isotropic, valley-degenerate energy dispersion. The trajectories exhibit symmetric Veselago focusing patterns~\cite{Cheianov2007} with diamond-shaped caustics between the two barrier interfaces, analogous to flat-lens focusing in a negative-refractive-index medium.

In borophene with $\alpha = 0^\circ$ [Fig.~3(b)], the tilt breaks the angular symmetry of the carrier trajectories. The $K$ carriers (blue) and $K'$ carriers (red) refract in different transverse directions upon entering the barrier, which gives rise to an asymmetric Veselago focusing pattern — a tilted analog of the graphene Veselago lens~\cite{Cheianov2007,ZhangPeeters2018}. The $K$-valley trajectories are deflected toward one transverse direction ($+y$), whereas the $K'$ trajectories are deflected toward the opposite direction ($-y$). However, despite this valley-dependent refraction, the straight barrier does not produce net valley filtering, since both valleys transmit with equal total weight when integrated over all incident angles. This is consistent with the zero-valley polarization shown in Fig.~2(a).

The picture changes qualitatively in Fig.~3(c), where the barrier is tilted by $\alpha = 20^\circ$. The angled barrier (indicated by the tilted black lines) reorients the refraction geometry such that $K$-valley carriers (blue) encounter the barrier near the oblique Klein tunneling peak and transmit more efficiently, producing dense blue trajectories on the drain side. In contrast, $K'$ carriers (red) approach the tilted barrier at unfavorable angles, are predominantly reflected, and appear as faint red trajectories confined to the source side. This spatial separation of the valley-resolved carrier flow clearly demonstrates the valley filtering mechanism: the tilted barrier preferentially passes through $K$-valley carriers while blocking the $K'$ valley.

\begin{figure}[H]
\centering
\includegraphics[width=0.95\textwidth]{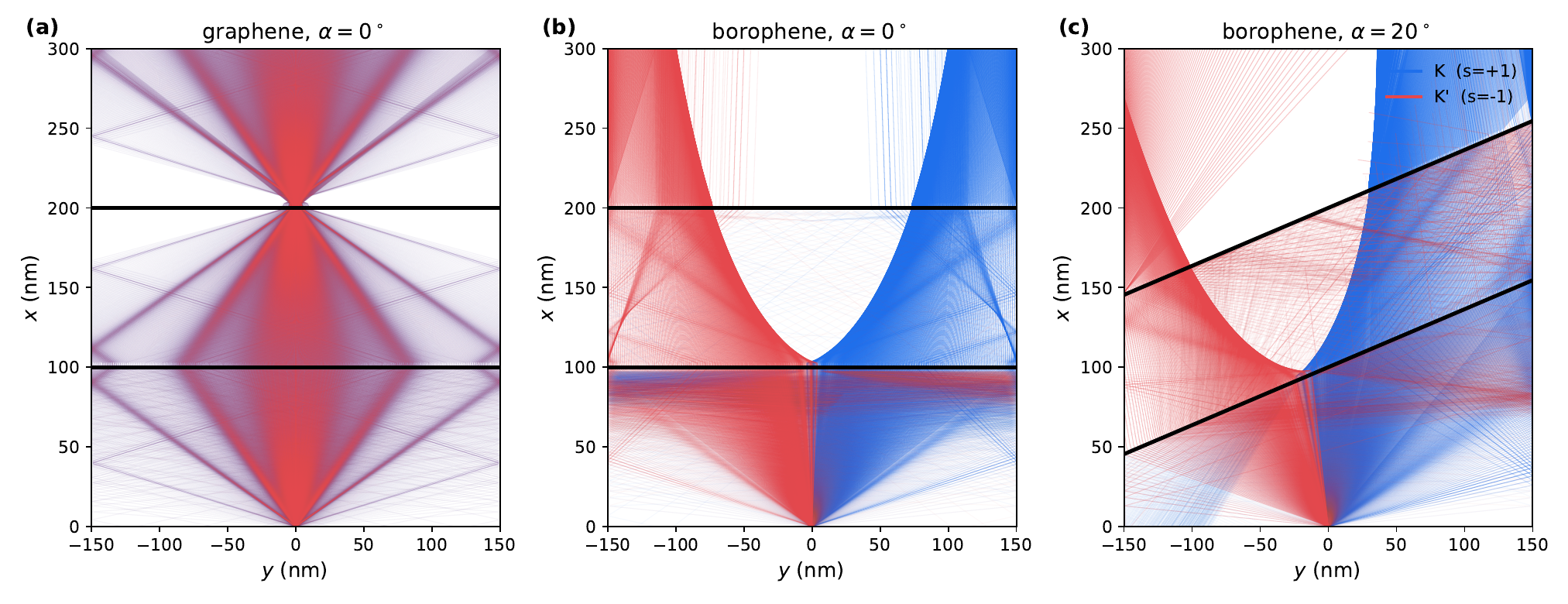}
\caption{\label{fig:trajectories}\textbf{Semiclassical carrier trajectories from valley degeneracy to valley filtering.} Each trajectory opacity is proportional to its transmission probability; blue: $K$ ($s = +1$), red: $K'$ ($s = -1$). Black lines indicate barrier boundaries. (a)~Graphene, $\alpha = 0^\circ$: degenerate $K$ and $K'$ trajectories with symmetric Veselago focusing. (b)~Borophene, $\alpha = 0^\circ$: the tilt induces valley-dependent refraction with asymmetric caustic patterns, but no net valley filtering. (c)~Borophene, $\alpha = 20^\circ$: the tilted barrier selectively transmits $K$ (blue) while reflecting $K'$ (red), clearly demonstrating the valley filtering effect. Parameters: $\vx/\vF = 0.86$, $\vy/\vF = 0.69$, $|\wt|/\vF = 0.32$, $\phi_{\mathrm{t}} = 90^\circ$ (8-\textit{Pmmn} borophene), $W = 300$~nm, $L = 300$~nm, $d = 100$~nm, $\EF = 80$~meV, $V_0 = 160$~meV.}
\end{figure}

\subsection{Physical mechanism of the valley filter}
\label{sec:mechanism}

The valley filtering mechanism can be understood as follows. Oblique Klein tunneling in tilted Dirac materials places the perfect-transmission directions at $+\theta_{\mathrm{K}}$ and $-\theta_{\mathrm{K}}$ for the $K$ and $K'$ valleys, respectively~\cite{Zhang2018,Kong2021}. For a straight barrier ($\alpha = 0$), the transmission profiles of the two valleys are mirror-symmetric of each other with respect to normal incidence. When integrated over all the transverse modes (i.e., over all $k_y$), the total conductance is identical for both valleys, resulting in $P = 0$. This is the situation depicted in Fig.~1(c) and is confirmed by the valley-degenerate curves in Fig.~2(a).

When the barrier is tilted by an angle $\alpha$, the conserved quantity changes from $k_y$ to $\kpar = k_x\sin\alpha + k_y\cos\alpha$, which effectively shifts the angular window through which carriers encounter the barrier. Crucially, this shift is the same for both valleys, however the two valleys have their Klein tunneling peaks at opposite angles ($\pm\theta_{\mathrm{K}}$), and the valley-independent angular shift produces a valley-dependent change in the integrated transmission: one valley is pushed closer to its Klein tunneling peak (enhancing total transmission), while the other is pushed away (suppressing it). The filtering efficiency depends on $\alpha$: as $\alpha$ increases from zero, it progressively breaks the valley symmetry by shifting the angular window through which each valley encounters the barrier, consequently enhancing the transmission of one valley relative to the other.

In contrast to previous valley filtering methods that utilize magnetic barriers~\cite{Settnes2016,Fujita2010,Yesilyurt2015,Yesilyurt2016AIP,Yesilyurt2019} or strain engineering~\cite{Guinea2010,Levy2010,Pereira2009}, the mechanism presented here is entirely electrostatic. The barrier angle $\alpha$ is fixed by the fabricated top-gate geometry, while the barrier height $V_0$ is tunable via the gate voltage. The sign of the valley polarization is governed by the sign of $\alpha$ (i.e., $P(\alpha) = -P(-\alpha)$), thereby enabling geometric control over the preferred valley. The magnitude of polarization increases with $V_0$ in the $n$-$p$-$n$ regime, as illustrated in Fig.~2(b,c), which allows for continuous electrical tuning.

\begin{figure}[H]
\centering
\includegraphics[width=0.95\textwidth]{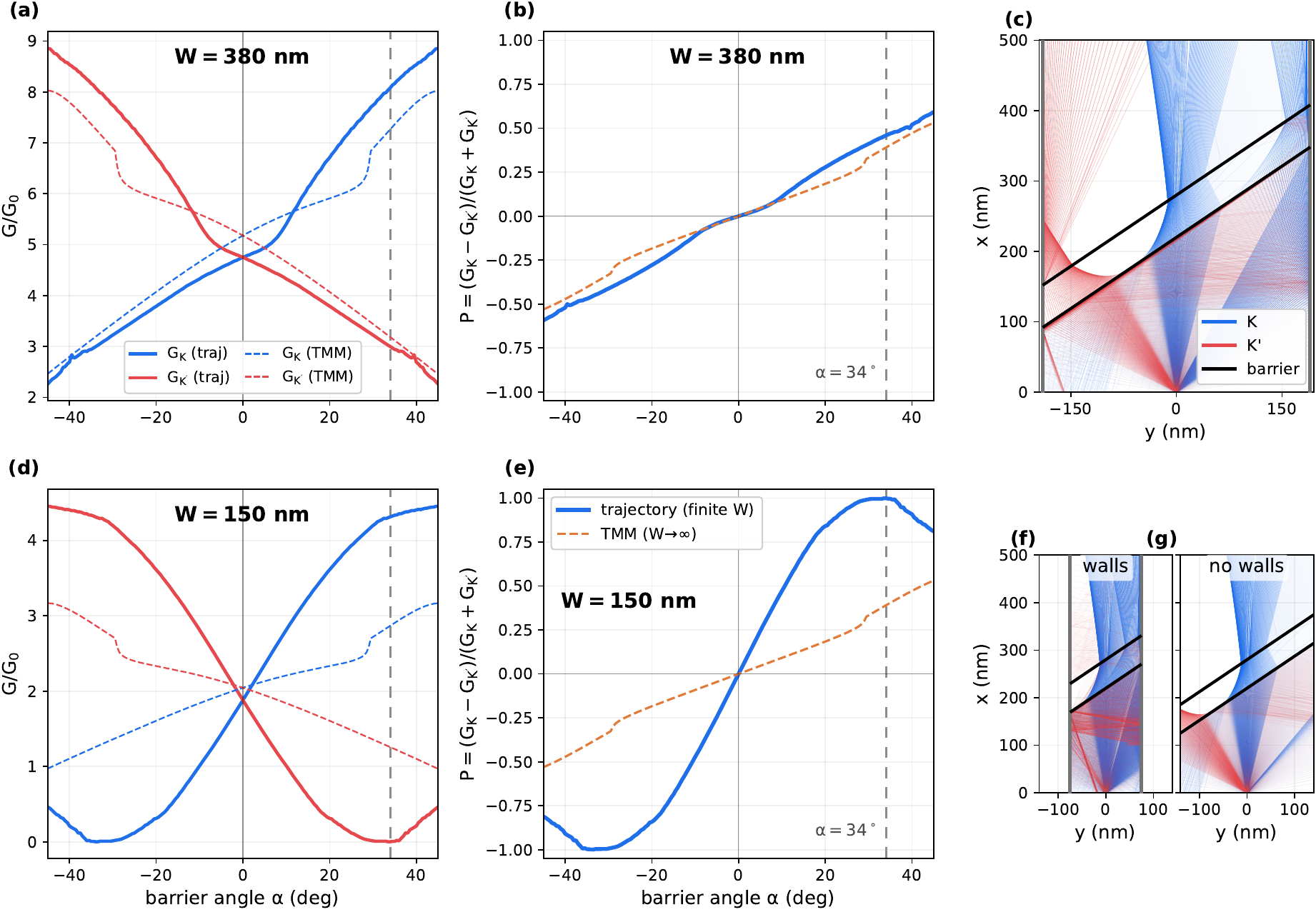}
\caption{\label{fig:mechanism}\textbf{Barrier-angle dependence of the valley filtering for a wide and a narrow channel.} The top and bottom rows correspond to the two channel widths $W$ indicated in the panels. (a),~(d)~Valley-resolved conductance $G/G_0$ as a function of barrier angle $\alpha$ for the $K$ (blue) and $K'$ (red) valleys, obtained from the semiclassical trajectory simulation (solid) and the transfer-matrix method (dashed). (b),~(e)~Valley polarization $P = (G_K - G_{K'})/(G_K + G_{K'})$ as a function of $\alpha$, obtained from the finite-width trajectory simulation (solid) and the plane-wave ($W \to \infty$) transfer matrix method (dashed); the vertical dashed line marks the barrier angle used in the trajectory panels. (c)~Real-space carrier trajectories for the wide channel. (f),~(g)~Real-space carrier trajectories for the narrow channel, with (f) and without (g) specular channel walls; blue: $K$, red: $K'$, black lines: barrier edges. Parameters: $\vx/\vF = 0.86$, $\vy/\vF = 0.69$, $|\wt|/\vF = 0.32$, $\phi_{\mathrm{t}} = 90^\circ$ (8-\textit{Pmmn} borophene), $\EF = 80$~meV, $V_0 = 120$~meV, $d = 60$~nm, the distance between the source and the first barrier interface $x_0 = 220$~nm, $L = 500$~nm, with channel widths $W = 380$~nm (top row) and $W = 150$~nm (bottom row); the real-space trajectory panels (c,~f,~g) use $\alpha = 34^\circ$.}

\end{figure}

This $\alpha$ dependence is quantified by the barrier-angle scan in Fig.~\ref{fig:mechanism}, where the dashed curves are the plane-wave TMM (infinite-width) result and the solid curves are the finite-width trajectory result using the same TMM transmission weight for each barrier encounter. The wide-channel result follows the TMM trend, while the narrow-channel geometry can strongly amplify the valley selectivity: near the marked design angle $\alpha\approx34^\circ$, wall-mediated recycling and finite-aperture selection drive the trajectory polarization close to $P\approx1$ over a finite angular interval. The smooth variation of $P(\alpha)$ also gives an estimate of the fabrication tolerance around the design angle.

\section{Methodologies and Possible Experimental Setups}
\label{sec:discussion}

\subsection{Transfer matrix formalism}
\label{sec:tmm_detail}

The coherent transmission $T(E, \kpar; s)$ is obtained from the transfer-matrix formalism. In a region of constant potential $V$, the tilt-shifted energy is $\tilde{\varepsilon} = (E - V) - \hbar s\wy k_y$, and the longitudinal wave vector satisfies

\begin{equation}\label{eq:kx_quad}
\bigl[(\hbar\vx)^2 - (\hbar s\wx)^2\bigr]k_x^2 + 2\tilde{\varepsilon}\,\hbar s\wx k_x + (\hbar\vy k_y)^2 - \tilde{\varepsilon}^2 = 0.
\end{equation}

For pure transverse tilt ($\wx=0$), this is reduced to $k_x^2 = [\tilde{\varepsilon}^2 - (\hbar\vy k_y)^2]/(\hbar\vx)^2$. The wavefunction matrix in each region is written as follows

\begin{equation}\label{eq:Phi}
\Phi(x) =
\begin{pmatrix}
\mathrm{e}^{\mathrm{i}k_x x} & \mathrm{e}^{-\mathrm{i}k_x x} \\
\lambda\chi_+\mathrm{e}^{\mathrm{i}k_x x} & \lambda\chi_-\mathrm{e}^{-\mathrm{i}k_x x}
\end{pmatrix},
\end{equation}

where $\lambda = \mathrm{sign}(\tilde{\varepsilon})$ and $\chi_\pm=(\pm\vx k_x+\mathrm{i}\vy k_y)/\sqrt{(\vx k_x)^2+(\vy k_y)^2}$. By matching the wavefunction at the two barrier interfaces, the total transfer matrix becomes

\begin{equation}\label{eq:M_total}
M_{\mathrm{total}} = \Phi_{\mathrm{free}}^{-1}(0)\Phi_{\mathrm{bar}}(0)\Phi_{\mathrm{bar}}^{-1}(\deff)\Phi_{\mathrm{free}}(\deff),
\end{equation}

and the transmission probability is given by $T = 1/|[M_{\mathrm{total}}]_{00}|^2$.

For $\alpha \neq 0$, the barrier-normal wave-vector $\kperp$ satisfies 

\begin{equation}\label{eq:kperp_quad}
  A_q\,\kperp^2 + B_q\,\kperp + C_q = 0\,,
\end{equation}

 with coefficients 

\begin{align}
  A_q &= \gamma^2 - (\hbar\vx)^2\cos^2\!\alpha - (\hbar\vy)^2\sin^2\!\alpha\,,\label{eq:Aq}\\
  B_q &= -2\varepsilon_{\mathrm{eff}}\gamma - 2\sin\alpha\cos\alpha\bigl[(\hbar\vx)^2 - (\hbar\vy)^2\bigr]\kpar\,,\label{eq:Bq}\\
  C_q &= \varepsilon_{\mathrm{eff}}^2 - \bigl[(\hbar\vx)^2\sin^2\!\alpha + (\hbar\vy)^2\cos^2\!\alpha\bigr]\kpar^2\,,\label{eq:Cq}
\end{align}

where $\gamma = \hbar\wx s\cos\alpha - \hbar\wy s\sin\alpha$ and $\varepsilon_{\mathrm{eff}} = (E - V) - \hbar\wx s\,\kpar\sin\alpha - \hbar\wy s\,\kpar\cos\alpha$. The physically correct root requires a positive group velocity along the barrier normal; a negative discriminant indicates total internal reflection. Note that these coefficients reduce to the standard expressions for the straight barrier ($\alpha = 0$) case~\cite{Kong2021,Zhang2018}. The key feature of Eqs.~\eqref{eq:Aq}--\eqref{eq:Cq} is that the barrier angle $\alpha$ couples to the tilt through $\gamma$ and to the anisotropy through the $\sin\alpha\cos\alpha$ cross-terms, which gives rise to the valley-dependent filtering mechanism described in this work.

\subsection{Semiclassical equations of motion}
\label{sec:semiclassical_detail}

In this approach, the sharp rectangular barrier is replaced by the smooth potential profile
\begin{equation}\label{eq:barrier_smooth}
V(\xeff) = \frac{V_0}{2}\Bigl[\tanh\!\Bigl(\frac{\xeff}{\sigma}\Bigr) - \tanh\!\Bigl(\frac{\xeff - d}{\sigma}\Bigr)\Bigr],
\end{equation}
where $\sigma$ is a small smoothing length that regularizes the step-function discontinuity and provides a continuous force gradient for numerical integration. The smoothing length is chosen to be much smaller than all physical length scales ($\sigma \ll d$), so that the potential closely approximates the sharp barrier used in the transfer-matrix calculation. Carriers are injected from the source and propagated through the device using the semiclassical equations of motion 
\begin{align}
  \dot{x} &= s\wx + \lambda\,\vx^2 k_x/\tilde{k}\,,\label{eq:xdot}\\
  \dot{y} &= s\wy + \lambda\,\vy^2 k_y/\tilde{k}\,,\label{eq:ydot}\\
  \hbar\dot{k}_x &= -\partial V/\partial x\,,\quad
  \hbar\dot{k}_y = -\partial V/\partial y\,,\label{eq:kdot}
\end{align} 
where $\tilde{k} = \sqrt{(\vx k_x)^2 + (\vy k_y)^2}$. The equations of motion are integrated using a fourth-order Runge--Kutta method with time step $\Delta t = 0.02$~fs. The smooth potential in Eq.~\eqref{eq:barrier_smooth} provides the continuous forces $-\partial V/\partial x$ and $-\partial V/\partial y = \tan\alpha\,\mathrm{d}V/\mathrm{d}\xeff$ that steer the carrier momentum as it traverses the barrier. Specular reflection is imposed at the channel walls ($y = \pm W/2$).

\subsection{Experimental prospects}
\label{sec:experimental}

The valley filtering demonstrated here relies on two generic ingredients: (i) a band anisotropy that renders the transmission valley-dependent in angle-space, thereby lifting the valley degeneracy, and (ii) an angled electrostatic barrier that converts this angular asymmetry into a net valley-polarized conductance. The framework is therefore not specific to a single material.

From a practical standpoint, WTe$_2$ is a promising platform for exploring valley-dependent transport properties. Although WTe$_2$ is a type-II Weyl semimetal, its dispersion has been reported to no longer be overtilted, i.e., tilted but type-I-like at approximately $130$~K~\cite{Lv2017}.

A concrete experimental realization can be envisioned: a WTe$_2$ nanoflake device with source and drain electrodes is equipped with an angled top gate that creates an $n$-$p$-$n$ barrier tilted by angle $\alpha$ relative to the current flow direction. Because the tilted Dirac cones occupy a finite energy window at a specific position in the band structure, operation requires the Fermi level to lie within that window; in the proposed device, this is set electrostatically by the global back gate, with chemical doping or substrate charge transfer as equivalent routes. Valley polarization could be detected through non-local voltage measurements or through the valley-dependent Fabry--P\'erot oscillation pattern in $G(V_0)$, which provides a distinctive spectroscopic fingerprint of valley filtering. Because $\alpha$ is fixed in a fabricated device rather than tuned in situ, several effective angles can be tested by fabricating gate arrays with different orientations, or by using a multi-terminal circular/polygonal device in which different source--drain pairs probe a fixed top-gate barrier from different transport directions. The crystal orientation and Dirac-cone tilt axis can be identified by anisotropic transport, polarized Raman, or diffraction measurements, and the chosen edge shape and contact apertures should be included in the corresponding finite-geometry simulation.

Beyond WTe$_2$, the organic conductor $\alpha$-(BEDT-TTF)$_2$I$_3$ has a stronger tilt, which may yield even stronger valley polarization. The present calculations assume ballistic transport and a sharp rectangular barrier profile; in practice, disordered scattering, finite-temperature phonon contributions, and smooth gate-defined potentials may modify the quantitative magnitude of the valley polarization. However, the underlying mechanism, i.e., the interplay between oblique Klein tunneling and the geometric symmetry breaking induced by the angled barrier, is robust to these effects, as it relies on the oppositely tilted Fermi-surface topology.

\section{Conclusion}
\label{sec:conclusion}

In summary, we showed that an angled electrostatic gate on a tilted Dirac material provides a purely electrostatic route to valley filtering in $n$-$p$-$n$ junctions, without the need for magnetic fields or strain engineering. We developed a generalized transfer-matrix formalism for the tilted, anisotropic Dirac Hamiltonian, extended it to barriers at arbitrary angles $\alpha$ relative to the transport direction, and derived the corresponding refraction equation in the rotated frame. The valley-resolved Landauer--B\"uttiker conductance of 8-\textit{Pmmn} borophene $n$-$p$-$n$ junctions shows that the $K$ and $K'$ conductances are degenerate for $\alpha=0$, whereas a clear valley-dependent conductance splitting emerges for $\alpha\neq0$ in the $n$-$p$-$n$ regime. Moreover, $P=(G_K-G_{K'})/(G_K+G_{K'})$ is highly dependent on the barrier angle and can approach unity for selected finite-width geometries. A semiclassical trajectory simulation in the full 2D device geometry directly shows how the angled barrier filters valleys: valley-dependent tunneling supplies the angularly asymmetric transmission kernel, while the gate orientation and finite channel geometry supply the required angular selection. The device parameters can be further optimized to achieve higher conductance and higher valley polarization, and the framework may pave the way for experimental realization of all-electrostatic valley filters in the growing family of materials hosting tilted Dirac and Weyl cones.

\section{Acknowledgments}

The author acknowledges the support of the Nanoelectronics Research Center (Nano Electronics Research Mühendislik Ar. Ge. Ltd. Şti., Istanbul, Turkey).

\section{Data availability}

The datasets generated and analyzed during this study are available from the corresponding author upon reasonable request; the simulation code used to generate them is openly available, as stated in the Code Availability Section.

\section{Code availability}

The transport simulator used in this work (transfer-matrix and semiclassical trajectory engines) is openly available under the MIT license at https://doi.org/10.5281/zenodo.20801655.

\section{Author contributions}

C.Y.\ conceived the idea, developed the theoretical framework, performed all calculations, and wrote the manuscript.

\section{Competing interests}

The author declares no competing interests.


\end{document}